# Stressing the Boundaries of Mobile Accessibility

Hugo Nicolau
Technical University of Lisbon / INESC-ID
Rua Alves Redol, 9
1000-029, Lisboa, Portugal
hman@vimmi.inesc-id.pt

João Guerreiro
Technical University of Lisbon / INESC-ID
Rua Alves Redol, 9
1000-029, Lisboa, Portugal
joao.p.guerreiro@ist.utl.pt

Tiago Guerreiro
LaSIGE, Department of Informatics, University of Lisbon
Campo Grande, C6 Piso3, 1749-016, Lisboa, Portugal
tjvg@di.fc.ul.pt

## ABSTRACT
Mobile devices gather the communication capabilities as no other gadget. Plus, they now comprise a wider set of applications while still maintaining reduced size and weight. They have started to include accessibility features that enable the inclusion of disabled people. However, these inclusive efforts still fall short considering the possibilities of such devices. This is mainly due to the lack of interoperability and extensibility of current mobile operating systems (OS). In this paper, we present a case study of a multi-impaired person where access to basic mobile applications was provided in an applicational basis. We outline the main flaws in current mobile OS and suggest how these could further empower developers to provide accessibility components. These could then be compounded to provide system-wide inclusion to a wider range of (multi)-impairments.

## Categories and Subject Descriptors
H.5.m. Information interfaces and presentation (e.g., HCI): Miscellaneous.

## 1. INTRODUCTION
Mobile devices are currently one of the most important tools for creating and maintaining social links. They comprise a large set of applications and functionalities that make them the ultimate communication tool, always within reach. The inability to control such devices is likely to exclude people from opportunities in several domains: work, entertainment, healthcare, shopping, transportation, and so forth.

These devices are expected to work in wide demographics, independently of social or economical status, age, preferences, values, or culture [1]. The diversity of their target audience is enormous and each individual has a very different set of requirements. However, current mobile interfaces do not address this need well. For instance, older adults may require larger targets and font size, due to increased physiological tremor and visual impairment. Auditory feedback and new touch-based exploration mechanisms are required for blind people. On the other hand, motor-impaired users may prefer voice interaction or alternative interaction styles rather than gesture and direct manipulation. All in all, mobile interfaces need to address a wide range of abilities by supporting parameterizations and adaptations, allowing its end-users to fully control their devices.

In the past two decades, desktop Operating Systems (OS) have evolved to support these needs by providing a set of accessibility features for hearing-impaired (e.g. ShowSounds[1]), motor-impaired (e.g. StickKeys[2]), and visual-impaired (e.g. VoiceOver[3]) people. As a result, researchers built solutions to automatically provide suggestions [4] or adaptations [5] for each user.

Although efforts have been made by most mobile OS[4,5], they fall short on the needs of mobile users. Compared to desktop computing, mobile accessibility is still in its infancy. In this paper, we describe a case study that illustrates the open challenges of mobile accessibility and discuss the limitations of the current mobile OS architecture towards a more inclusive development.

## 2. A Multi-Impairment Case-Study
The difficulties impaired users face when dealing with mobile devices are exacerbated for people with multiple impairments. We came across Michael, a 35 year old user, eager for social interaction. Due to an accident at the age of 21, Michael is tetraplegic. He only has residual arm and neck movement. Furthermore, the accident led to both blindness and a speech impairment that makes him stutter.

Michael is unable to autonomously control his mobile device, even with existing accessibility solutions. Blindness and tetraplegia preclude him from target discrimination and, therefore, from selecting options in both keypad and touch-screen devices. Voice interaction is hampered by his speech impairment and poses several privacy issues. These limitations block rather simple tasks, such as placing and answering phone calls. Michael's arm residual movement, when somehow supported, allows him to hit an indiscriminate area at his hand's range. The low fine motor control and lack of strength require a large and sensitive input device, such as a switch.

In order to provide Michael control over his mobile device, we developed an *android* application that replaces the devices' interface with a much simpler one. It scans through a set of menu options via audio and resorts to a switch button to select the current option. Our swift and low cost solution was based on a *mouse*, which hardware was modified to produce the same (unique) event for both buttons. This mouse was connected to a *Samsung Galaxy X* via *Bluetooth* and each click represented a tap on the screen. The supported features include the *Clock*, *Battery*, *Make Call* (favorite contacts at first, while the others are divided in groups – alphabetically), *Missed Calls* and *Messages Received*.

---

[1] http://www.microsoft.com/resources/documentation/windows/xp/all/proddocs/en-us/access_showsounds_turnon.mspx?mfr=true

[2] http://windows.microsoft.com/en-US/windows-xp/help/using-stickykeys

[3] http://www.apple.com/accessibility/voiceover/

[4] http://www.apple.com/accessibility/

[5] http://developer.android.com/design/patterns/accessibility.html

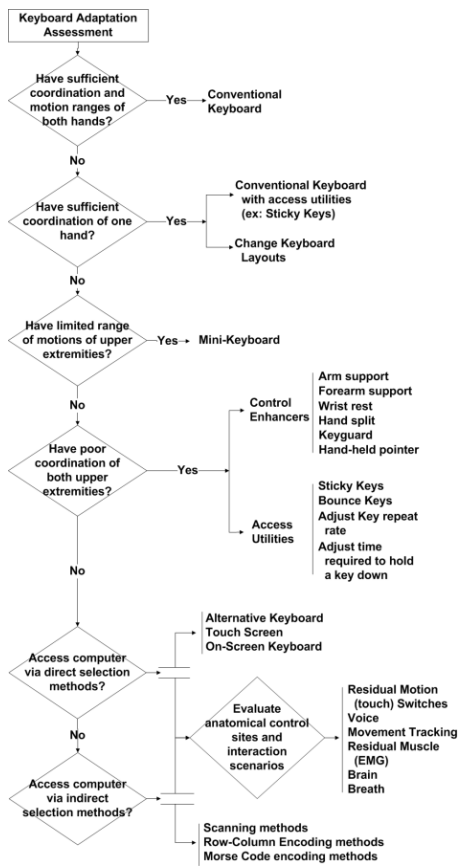

**Figure 1 - Keyboard Adaptation Assessment Tree for Motor Impairments [6]**

All menus are read twice, giving the user opportunity to select an option in case he missed it. Also, a "*Back*" option is present at the beginning of all menus. This option showed to be of upmost importance to be located in the beginning of all menus in order to deal with undesired mistakes.

Michael has been using this prototype for about two months. He is now able to autonomously make and answer to phone calls and talk to his friends and family. Michael's next requirement regards entering text to write SMSs, to make phone calls by inserting the phone number and to enter new contacts. However, what these and future requirements portray is the inadequacy of current OS and interfaces to support people with such impairments. With the design of specific applications we may enhance a person's life (or of a specific population), yet, functionality is restricted to what each application conveys. Alternatively, mobile devices should provide "system-wide" control mechanisms to allow people with different kinds of impairments to enjoy the fullness of their abilities.

## 3. Mobile Accessibility Panorama

The emergence of built-in accessibility features along with device size and communication abilities is presenting the mobile phone as a tool towards inclusion. As an example, several blind people acquired the desired mobile access with the advent of Apple's iPhone and VoiceOver. Others came along. Meanwhile, as stressed in the presented case study, mobile devices are still a challenge for several people.

One important aspect is that the accessibility features provided by current mobile OS are strictly prepared for single impairments and, looking at it close enough, only address a subset of disabilities tackled by desktop-based assistive technologies. Looking at screen readers, for instance, they allow blind people to interact with a mobile device but obligate them to a similar physical action as a sighted person. When blindness comes along with physical impairments as in the case of Michael, access to mobile applications becomes compromised. One reason for this limited accessibility is that current mobile OS lack the flexibility to support accessible compounded and integrated development. Making the analogy with desktop-based assistive technologies, these are often created and used by merging different components (applications and device drivers) thus providing the ability to tackle needs that go beyond the stereotypical single-impairment. Figure 1 shows a keyboard adaptation assessment tree strictly focusing on motor impairments where from the point where the user shows poor coordination on both upper extremities solutions are offered via the conjunction of both hardware peripherals and software adaptations. The latter are often deployed to emulate keyboard and mouse events which enable system-wide usage and thus foster the seamless inclusion of disabled people. If we step up to a multi-impairment scenario, different components (once again system-wise) can be put together to empower their users.

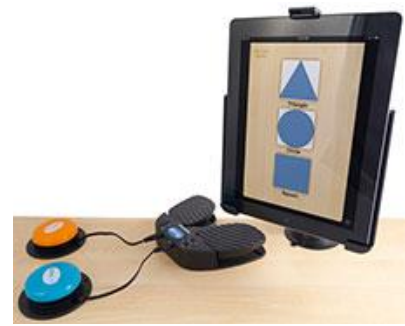

**Figure 2 - Blue2 Bluetooth Switch by AbleNet®**

On the mobile side, system-wide assistive technology is restricted to the one developed by manufacturers. The remaining components are developed or supported at application level. Figure 2 presents a switch enabling access to iPhone/iPad applications but only to those that were designed to support it (e.g., SoundingBoard AAC app[6]). Several others work similarly. The lack of interoperability between applications and the OS is also patent in softer-level adaptations: once again, current mobile OS lack the adaptability features to address the needs of a dynamic and varied population. Recent studies have presented results to support the adaptability and personalization of mobile interfaces [2, 3]. However, applying these adaptations system-wide is still very difficult or unfeasible as these prominent interfacing decisions are hard-coded and unreachable to the user or application developer.

We argue that mobile devices (i.e., their mobile OS) should enable higher degrees of interoperability pertaining all aspects of interfacing that relate to output and input. Once again, making the analogy with desktop computers, it should be possible to:

- Offer alternative feedback for current focus/selection;

---

[6] http://www.ablenetinc.com/AssistiveTechnology/Communication/SoundingBoard/tabid/632/Default.aspx

- Have control over selection method, enabling navigation between items (e.g. simple or directed scanning);
- Filter and adapt user input;
- Parameterize and adapt interfacing restrictions (e.g., timeouts) to fit a particular individual
- Parameterize rendering attributes (e.g., prepare output image for particular users and disabilities; color-blind)

All this should be provided system-wide so that disabled people can seek to attain similar access and opportunities as non-disabled people.

## 4. ACKNOWLEDGMENTS
Work supported by the Portuguese Foundation for Science and Technology (FCT): individual grant SFRH/ BD/66550/2009; project PAELife AAL/0014/2009; and project PEst-OE/EEI/LA0021/2011.

## 5. REFERENCES
[1] Budde, P. *Global Mobile Communications – Statistics, Trends and Forecasts*. Tech. Report (2009).

[2] Montague et al. Designing for Individuals: Usable Touch-Screen Interaction through Shared User Models. In Proc. ASSETS'12, 151-158.

[3] Oliveira et al. Blind People and Mobile Touch-based Text-Entry: Acknowledging the Need for Different Flavors. In Proc. of ASSETS'11 (2011), 179-186.

[4] Trewin, S. *Automating Accessibility: The Dynamic Keyboard*. In Proc. of ASSETS'03 (2003), 71-79.

[5] Wobbrock, J. et al. The Angle Mouse: Target-Agnostic Dynamic Gain Adjustment Based on Angular Deviation. In Proc. of CHI'09 (2009), 1041-1410.

[6] Wu, Ting-Fang, et al. "Computer access assessment for persons with physical disabilities: A guide to assistive technology interventions." *Computers helping people with special needs* (2002): 195-214.